\documentclass[a4paper,11pt]{article}
\pdfoutput=1
\usepackage{latexsym}
\usepackage{amsmath}
\usepackage{amsfonts}
\usepackage{graphicx}

\def\laq{~\raise 0.4ex\hbox{$<$}\kern -0.8em\lower 0.62
ex\hbox{$\sim$}~}
\def\gaq{~\raise 0.4ex\hbox{$>$}\kern -0.7em\lower 0.62
ex\hbox{$\sim$}~}

\def\beq{\begin{equation}}
\def\eeq{\end{equation}}
\def\bea{\begin{eqnarray}}
\def\eea{\end{eqnarray}}
\def\bean{\begin{eqnarray*}}
\def\eean{\end{eqnarray*}}

\def\re {\rangle}
\def\eff{e\!f\!f}

\def \pa {\partial}

\def \ti {\widetilde}

\def \a {\alpha}

\def \ep {\epsilon}

   \def\be{\begin{equation}}
   \def\ee{\end{equation}}
   \def\ba{\begin{eqnarray}}
   \def\ea{\end{eqnarray}}

\title{Cosmological Backreaction for a Test Field Observer in a Chaotic Inflationary Model}

\author{
 \ 
Giovanni Marozzi$^{1,2}$\thanks{giovanni.marozzi@unige.ch} \ ,
             Gian Paolo Vacca$^3$\thanks{vacca@bo.infn.it} \, ,
             Robert H. Brandenberger$^4$\thanks{rhb@physics.mcgill.ca} \\
      \\ 
      ${}^1$  Coll\`ege de France, 11 Place M. Berthelot, 75005 Paris, France.\\
      ${}^2$  Universit\'e de Gen\`eve, D\'epartement de Physique Th\'eorique and CAP,\\
24 quai Ernest-Ansermet, CH-1211 Gen\`eve 4, Switzerland. \\
      ${}^3$ INFN Sezione di Bologna, 
      via Irnerio 46, I-40126 Bologna, Italy.\\    
      ${}^4$ Physics Department, McGill University, \\
3600 University Street, Montreal, QC H3A 2T8, Canada  
}
\date{}
\begin{document}
\addtolength{\belowdisplayskip}{-0.0cm}       
\addtolength{\abovedisplayskip}{-0.0cm}       

\maketitle
\begin{abstract} 

In an inhomogeneous universe, an observer associated with a particular matter field does not
necessarily measure the same cosmological evolution as an observer in a homogeneous and
isotropic universe. Here we consider, in the context of a chaotic inflationary background model,
a class of observers associated with a ``clock field" for which we use a light test field. We
compute the effective expansion rate and fluid equation of state in a gauge invariant way, 
taking into account the quantum fluctuations of the long wavelength modes, and working up 
to second order in perturbation theory and in the slow-roll approximation. 
We find that the effective expansion rate is smaller than what
would be measured in the absence of fluctuations. Within the stochastic approach we study the 
bounds for which the approximations we make are consistent. 

\end{abstract}

\section{Introduction}

The setting of this work is inflationary cosmology \cite{Guth}, specifically a simple large
field (or ``chaotic") inflationary model \cite{Linde} in which the almost constant
potential of a slowly rolling scalar field $\Phi$ (the ``inflaton") leads to a phase of almost exponential
expansion of space. We know, however, that there must be other matter fields
present, in particular those of the Standard Model of particle physics. These fields
may play a sub-dominant role in determining the expansion of space during the
period of inflation, but they are very relevant to late-time cosmology. 
In such a framework it is reasonable to consider the possibility of associating to 
one of these extra fields a class of dynamical observers, a fact
which leads to the construction and study of observables which are determined by 
measurements performed by such a class of observers. In the presence of
primordial cosmological fluctuations, which are unavoidably present in the semiclassical 
approach of the gravity-matter dynamics, such observers will in general not measure exactly
the same physics as a canonical observer in a homogeneous universe would.
Such an effect is, in general, called the cosmological backreaction.

This is a non-trivial task and we shall consider here a simple toy model setup.
Following \cite{Ghazal2}, we will consider the case where the clock/observer is 
associated to a light test scalar field $\chi$ (as opposed to the inflaton field), 
and the inflaton field has the standard 
quadratic potential of chaotic inflation given by $V(\Phi)=\frac{m^2}{2} \Phi^2$. 
More specifically, we assume that the light field $\chi$
at the background level does not affect the evolution of the space-time during the 
inflationary regime but only plays the role of a clock. 

In general, the issue of backreaction of cosmological perturbations \cite{ABM} can
be described as follow.
If we start with a background cosmological solution (metric and matter) and add to it
small amplitude fluctuations of relative amplitude $\epsilon$ \footnote{Note that $\epsilon$ is
not the inflationary slow-roll parameter.} which satisfy the linear perturbation equations,
then due to the non-linearities in the Einstein equations these equations are not satisfied to
quadratic order. Second order fluctuations in matter and metric build up. In particular, corrections
to the background can be induced. The calculations of \cite{ABM} showed
that, to leading order in spatial gradients, super-Hubble linear fluctuation modes induce 
effects which grow  logarithmically in time and act like a negative contribution to the
effective cosmological constant. It was then conjectured \cite{RHBrev} that this effect
might lead to a dynamical relaxation mechanism for the cosmological constant,
in analogy of the effect which is conjectured to occur in pure quantum gravity due to
two loop infrared gravitons \cite{TW}. In \cite{ABM}, the effects were computed as a 
function of the background time variable $t$. 

However, in \cite{Unruh} the challenge
was raised as to whether this effect could be measured in terms of a physical
clock. Indeed, it was then shown \cite{Ghazal1} that in the case of single field
matter, the infrared backreaction  is not physically
measurable (see \cite{FMVVprl} for the same result obtained within a manifestly 
gauge invariant, and up to second order in perturbation theory, approach). 
The Hubble expansion rate measured by the only available
physical clock (namely the value of the scalar field) has the same history
with and without long wavelength fluctuations (see also \cite{Abramo}).

Now, in late time cosmology we measure time in terms of the temperature
of the cosmic microwave background (CMB) which at late times is a sub-dominant
component of matter. Hence, as said, it is reasonable to ask whether effects of
cosmological fluctuations can be measured by a clock which corresponds
to a sub-dominant field, a field which in the late time universe would
represent the temperature of the CMB. A first step in this direction was made
in \cite{Ghazal2} where it was shown that a clock tied to an entropy
field will indeed measure a different expansion rate of space in the
presence and absence of curvature fluctuations.

On the other hand, in order to give meaningful statements about observables - once the 
observers have been chosen - these must be constructed in a gauge invariant way, 
i.e. they should be defined covariantly in a reparameterization invariant way,
so that any other spectator (secondary) observer agrees with the definition given, 
computing it with its preferred reference system.
Therefore some care is required since normally one is dealing with observables 
obtained after averaging procedures. For example, it is crucial
to take into account the volume measure.
The problem can be naturally analyzed employing the gauge invariant method 
developed in~\cite{GMV1,GMV2}  (where the set of effective
equations for the averaged geometry given in  \cite{Buchert} has been
generalized in a covariant and gauge invariant form)
and recently applied to single field inflationary 
models~\cite{FMVVprl,MV}. 

Therefore, considering the gauge invariant approach just mentioned,
the question we address in this paper is the following: we compare
the average expansion rate of a spatial hypersurface characterized by
a fixed value of a spectator test scalar field in two space-time manifolds,
the first one which has no curvature fluctuations, the second one
which contains such perturbations~\footnote{Let us observe that, in investigations 
concerning the actual dynamics of the Universe, 
one may imagine more suitable choices to investigate physical phenomena 
based on past light-cone observations.
In particular, the gauge invariant approach of \cite{GMV1,GMV2} has been generalized 
to averaging over the past light-cone of a generic observer in \cite{LC} and this 
new approach has been applied in \cite{LCapplication}.}. 
We find that as a consequence of
long wavelength fluctuations, the average expansion rate is smaller.
This effect is at quadratic order in the amplitude of cosmological fluctuations
(at one loop in quantum field theory language) as opposed to
the situation in pure gravity where backreaction effects of long
wavelength fluctuations occur only at two loop order \cite{TW}.
The backreaction leads to a reduction of the expansion rate. Whether
the effect leads to a uniform relaxation of the cosmological constant
or simply to a stochastic adjustment is probably difficult to be answered 
using only our perturbative approach.

The outline of this paper is as follows. In Section 2 we
review the covariant and gauge-invariant approach which we
shall use, and introduce the ``effective Hubble expansion rate"
for an observer in an inhomogeneous universe. Section 3 forms the 
heart of this paper: we introduce our toy model, define the class of 
observers used (observers associated 
with a test scalar field $\chi$), determine the gauge transformation into 
a coordinate system in which $\chi$ is constant on spatial hypersurfaces,
and evaluate the effective Hubble expansion rate which our observer
sees. We also discuss bounds on the parameters of the model under which
our perturbative analysis is valid, estimate the magnitude of the
terms contributing to the backreaction, and determine the equation
of state as seen by our observers. In Section 4 we give a general discussion 
and provide some numerical results. The final section contains our
conclusions.
 
\section{Observables and gauge invariant backreaction}

The issue of the backreaction on an observable induced by quantum 
fluctuations in an inflationary universe can be cast on solid grounds 
using a recent covariant and gauge invariant (GI) approach~\cite{GMV1,GMV2}. 
One can construct a general non-local observable by taking
quantum averages of a scalar field $S(x)$ over a space-time hypersurface
$\Sigma_{A_0}$ where another scalar $A(x)$ (assumed to have a time-like gradient) 
has constant value $A_0$. Given this second scalar field, we can define a
particular (barred) coordinate system $\bar{x}^{\mu}=(\bar{t}, \vec{x})$ 
in which the scalar $A$ is homogeneous. A physical (and hence GI)
definition of the observable is then given by the following quantity
\beq
\langle S \rangle_{A_0}={\langle \sqrt{|\overline{\gamma}(t_0, {\vec{x}})|} 
\,~ \overline{S}(t_0, {\vec{x}}) \re \over  \langle   
\sqrt{|\overline{\gamma}(t_0, {\vec{x}})|} \re}\,,
\label{media}
\eeq
where we have denoted by $\sqrt{|\overline{\gamma}(t_0, {\vec{x}})|}$  the
determinant of the induced three dimensional metric on $\Sigma_{A_0}$.
The natural foliation of spacetime is then defined by the four vector 
\be
n^\mu= -\frac{\partial^\mu A}{(-\partial^\nu A\partial_\nu A)^{1/2}}
\ee
which characterizes the class of observers. 

We are interested in the dynamics encoded in the effective scale factor
$a_{eff}=\langle\sqrt{|\bar{\gamma}|}\, \rangle ^{1/3}$,
which satisfies the GI effective cosmological equation~\cite{GMV2} 
\be
\left(\frac{1}{a_{eff}}\frac{\partial \, a_{eff}}{\partial A_0} \right)^2
=\, \,\frac{1}{9} \left\langle\frac{\Theta}{(-\partial^\mu A\partial_\mu A)^{1/2}} 
\right\rangle_{A_0}^2\,,
\label{genEQ}
\ee
where  $\Theta=\nabla_\mu n^\mu$ is the expansion scalar of the timelike 
congruence $n^\mu$. This effective scale factor describes the expansion of 
the space as seen by the class of observers sitting on the hypersurface $\Sigma_{A_0}$.
The dynamics can be then extracted solving the Einstein and matter equations 
of motion in any gauge.

In order to deal with the metric components in any specific frame we employ
the standard decomposition of the metric in terms of scalar, transverse vector 
($B_i$,$\chi_i$) and traceless transverse tensor ($h_{ij}$) 
fluctuations up to the second order
around a homogeneous Friedmann-Lemaitre-Robertson-Walker (FLRW) zero order space-time
\bea
& & \,\,\,\,\,\,\, \,\,\,\,\,\,\,\,\,\,\,\,\,\,\,\,\,\,g_{00}= -1-2 \a-2 \a^{(2)}\,\,\,\,\,\,\,, \,\,\,\,\,\,\, 
g_{i0}=-{a\over2}\!\left(\beta_{,i}+B_i \right) 
-{a\over2}\left(\beta^{(2)}_{,i}+B^{(2)}_i\!\right) 
\nonumber\\
& & \!\!\!\!\!\!g_{ij}=  a^2 \!\Bigl[ \delta_{ij} \! 
\left( \!1\!-\!2 \psi\!-\!2 \psi^{(2)}\right)
+D_{ij} (E+E^{(2)})
+{1\over 2} \left(\chi_{i,j}+\chi_{j,i}+h_{ij}\right)
+ {1\over 2} \left(\chi^{(2)}_{i,j}+\chi^{(2)}_{j,i}+h^{(2)}_{ij}\right)\Bigr]
\label{GeneralGauge}
\eea
where $D_{ij}=\partial_i\partial_j-\delta_{ij}\nabla^2/3$ and for notational
simplicity we have removed a superscript for first order quantities.
The Einstein equations connect these fluctuations with those in the matter
fields.  For example, an inflaton field can be written to second order as
$\Phi(x)=\phi^{(0)}(t)+\phi^{(1)}(x)+\phi^{(2)}(x)$ (where $x$ is the space-time
four vector).

These general perturbed expressions can be gauge fixed. Let us recall
some common gauge fixings of the scalar and vector part:
the synchronous gauge (SG) is defined by $g_{00}=-1$ and $g_{i0}=0$,
the uniform field gauge (UFG) apart from setting $\Phi(x)=\phi^{(0)}(t)$ must be
supplemented by other conditions (one can consider $g_{i0}=0$),
finally the uniform curvature gauge (UCG) is defined by
$g_{ij}=a^2\left[\delta_{ij}+\frac{1}{2} \left(h_{ij}+h^{(2)}_{ij}\right)\right]$.

To quadratic order (in the amplitude of the fluctuations), the inhomogeneities
effect the effective expansion rate. This is called the ``backreaction" effect.
Using Eq.~\eqref{genEQ} we can now define the backreaction on the averaged 
expansion rate as the one obtained with respect to a class of observers identified 
by the scalar field $A(t, \vec{x})$.
In the long wavelength (LW) limit and neglecting the tensor perturbations, we 
obtain~\cite{FMVVprl}
\be
H^2_{eff}\equiv \dot{A}^{(0)\,2} \left(\frac{1}{a_{eff}}\frac{\partial \, a_{eff}}{\partial A_0} \right)^2 =
H^2 \left[1+\frac{2}{H}\langle \bar{\psi}\dot{\bar{\psi}}\rangle-
\frac{2}{H}\langle \dot{\bar {\psi}}^{(2)}\rangle\right]\,,
\label{EQ1simpl_2}
\ee
Let us stress here that the choice of a particular class of observers leads to the choice of a 
particular observable. The observers depend on a preferred reference frame but the 
associated observable is physical, and hence gauge invariant, in the sense that every 
other observer associated to a different frame agrees on its value.

In an inflationary background, the long wavelength contribution to the expectation 
values $\langle {\bar{\psi}}\dot{\bar{\psi}} \rangle$ and 
$\langle \dot{\bar {\psi}}^{(2)}\rangle$ will be increasing in time since the
phase space of infrared (super-Hubble) modes is growing. The contribution of
the ultraviolet (sub-Hubble) modes is constant in time since, given an ultraviolet
cutoff which is fixed in physical coordinates (see e.g. \cite{Wei} for a discussion
of this point), the phase space of these modes is constant (in the limit of exponential
inflation). In the following we will use the stochastic approach to evaluate the
above expectation values.

\section{Gauge invariant backreaction in a 2-field model}

The classical model from which we start is a Friedmann-Lemaitre-Robertson-Walker (FLRW) 
space-time with an inflaton (scalar) field $\Phi$ and a second light field $\chi$ which we 
study in the test field approximation (i.e. neglecting its energy density and pressure in the
background FLRW equations). 
We shall admit all possible inhomogeneous scalar quantum fluctuations of the metric and 
of the scalar matter fields, up to second order in perturbation theory.
In particular, the inflaton model we consider here is, for simplicity, given by a free massive 
scalar field which we shall treat in the slow-roll approximation together with a second free 
massive (test) field $\chi$, both minimally coupled to gravity. The action is
   \be
    S = \int d^4x \sqrt{-g} \left[ 
\frac{R}{16{\pi}G}
    - \frac{1}{2} g^{\mu \nu}
    \partial_{\mu} \Phi \partial_{\nu} \Phi
    - \frac{1}{2} m^2 \Phi ^2  - \frac{1}{2} g^{\mu \nu}
    \partial_{\mu} \chi \partial_{\nu} \chi-\frac{1}{2} m_\chi^2 \chi ^2 
\right] \,.
    \label{action}
    \ee
In the rest of this work we shall make an heavy use of the results on the stochastic approach
to inflationary dynamics presented in Section $5$ of~\cite{FMSVV2}.

\subsection{Perturbative dynamics of the observers}

We now choose a special class of observers which will define our observables.
Following \cite{Ghazal2}, we consider the average expansion rate as computed 
by the observers "$\chi$", namely by the observers sitting on the 
spacelike hypersurface defined by $\chi$ equal to a constant.  This hypersurface 
characterizes what we denote the barred gauge. 
To proceed we need the solutions of the equation of motion in this frame up to 
second order in perturbation theory (see Eq.~\eqref{EQ1simpl_2}).
These can be obtained with a gauge transformation from the results in the 
UCG~\cite{FMSVV2, FMVV2004}. As a consequence we have to find the gauge 
transformation that goes from the UCG to the barred gauge,
where  $\chi(x)=\chi^{(0)}(t)+\chi^{(1)}(x)+\chi^{(2)}(x)$ is equal to a constant, i.e.
where $\chi^{(1)}= \chi^{(2)}=0$.
To fix  the second gauge completely we also impose the condition $\beta=0$. 
This gauge, which we denote U$\chi$FG, explicitly depends on the property of the 
field $\chi$. We stress that this is a new scenario compared to previous analysis 
in a single field inflationary model~\cite{FMVVprl,MV}, where the observers were defined
in a geometrical way.

To find the gauge transformation, let us first consider the ``infinitesimal'' coordinate 
transformation parametrized by the  first-order and second-order
generators $\ep_{(1)}^\mu$ and  $\ep_{(2)}^\mu$, respectively,  and
given by \cite{MetAll}
\beq
x^\mu \rightarrow \tilde{x}^\mu= x^\mu + \epsilon^\mu_{(1)} +\frac{1}{2}
\left(\epsilon^{\nu}_{(1)}\pa_\nu \epsilon^{\mu}_{(1)} + \epsilon^{\mu}_{(2)}\right) + \dots
\label{311}
\eeq
where
\beq
\ep_{(1)}^\mu= \left( \ep_{(1)}^0, \pa^i \ep_{(1)}+ \ep_{(1)}^i 
\right), ~~~~~~~~~
\ep_{(2)}^\mu= \left( \ep_{(2)}^0, \pa^i \ep_{(2)}+ \ep_{(2)}^i \right) 
\label{312}
\eeq
(we have explicitly separated the scalar part from the pure transverse vector 
part $\ep_{(1)}^i$, $\ep_{(2)}^i$). Let us then recall that the associated gauge transformation 
of a scalar field $S$ is, to first order, 
\be
S^{(1)}~~ \rightarrow ~~\ti{S}^{(1)}=S^{(1)}-\ep_{(1)}^0 \dot{S}^{(0)},
\label{315}
\ee
and, to second order, 
\bea
S^{(2)} ~~\rightarrow &&~~\ti{S}^{(2)}=  S^{(2)}-\ep_{(1)}^0 \dot{S}^{(1)}
-\left(\ep_{(1)}^i+\partial^i \ep_{(1)}\right) \partial_i S^{(1)}
\nonumber \\ 
&&+\frac{1}{2}\left[\ep_{(1)}^0 
\partial_t (\ep_{(1)}^0 \dot{S}^{(0)})+ \left(\ep_{(1)}^i+\partial^i \ep_{(1)}\right)\partial_i
\ep_{(1)}^0 \dot{S}^{(0)}-\ep_{(2)}^0 \dot{S}^{(0)}\right]\,.
\label{316}
\eea 

On neglecting vector perturbations we can fix to zero $\ep_{(1)}^i$ and with 
$S\equiv \chi$ the gauge conditions $\tilde{\chi}^{(1)}=0$ and $\tilde{\chi}^{(2)}=0$ give 
the following results:
\be
\epsilon_{(1)}^0=\frac{\chi^{(1)}}{\dot{\chi}^{(0)}} \,\,,\,\,
\label{GT1}
\ee
\be 
\epsilon_{(2)}^0=2 \frac{\chi^{(2)}}{\dot{\chi}^{(0)}}+ \frac{1}{\dot{\chi}^{(0)}}
\left[-2 \left(\partial^i \ep_{(1)}\partial_i \chi^{(1)}+\ep_{(1)}^0 \dot{\chi}^{(1)}\right)
+\ep_{(1)}^0\partial_t\left(\ep_{(1)}^0 \dot{\chi}^{(0)}\right)
+\left(\partial^i \ep_{(1)}\partial_i \ep_{(1)}^0\right) \dot{\chi}^{(0)}\right] \, ,
\ee
and using $\epsilon_{(1)}^0$ in  $\epsilon_{(2)}^0$ one obtains
\be
\epsilon_{(2)}^0=2 \frac{\chi^{(2)}}{\dot{\chi}^{(0)}}-\frac{1}{\dot{\chi}^{(0)}}
\partial^i \ep_{(1)}\partial_i \chi^{(1)}-\frac{1}{\dot{\chi}^{(0)\,2}}\chi^{(1)}
\dot{\chi}^{(1)}\,.
\label{GT2}
\ee
We can now evaluate $\epsilon_{(1)}$ from the second gauge condition of U$\chi$FG,
namely $\tilde{\beta}=0$. One obtains  (from Eq. (3.8) of \cite{Marozzi})
\be
\epsilon_{(1)}=\int dt \left[\frac{1}{a^2}\frac{\chi^{(1)}}{\dot{\chi}^{(0)}}-\frac{\beta}{2 a}
\right]\,.
\label{GT3}
\ee

Considering finally the transformation between UCG and U$\chi$FG it is straightforward to 
derive the following gauge transformation for $\bar{\psi}$
and $\bar{\psi}^{(2)}$ (from Eqs. (3.9) and (3.13) of \cite{Marozzi}):
\be
\bar{\psi}=H \epsilon_{(1)}^0+\frac{1}{3} \nabla^2 \epsilon_{(1)}
\ee
\be
\bar{\psi}^{(2)}=\frac{H}{2} \epsilon_{(2)}^0 +\frac{1}{6} \nabla^2 \epsilon_{(2)}
-\frac{H}{2} \epsilon_{(1)}^0 \dot{\epsilon}_{(1)}^0 -\frac{\epsilon_{(1)}^{0\,2}}{2}
\left(\dot{H}+2 H^2 \right)-\frac{H}{2} \partial_i \epsilon_{(1)}^0
\partial^i  \epsilon_{(1)}-\frac{1}{6} \Pi^i_i\,,
\ee
where $\Pi_{i j}$ is defined in Eq. (3.16) of \cite{Marozzi}.
Using the above results and working in the long wavelength limit (and neglecting 
also tensor perturbations) one explicitly obtains
\be 
\bar{\psi}=\frac{H}{\dot{\chi}^{(0)}}\chi^{(1)}\,\,\,\,\,\,,\,\,\,\,\,\,
\bar{\psi}^{(2)}=\frac{H}{\dot{\chi}^{(0)}}\chi^{(2)}-\frac{H}{\dot{\chi}^{(0)\,2}}\chi^{(1)}
\dot{\chi}^{(1)} -\frac{\chi^{(1)\,2}}{2 \dot{\chi}^{(0)\,2}}
\left(2 H^2+\dot{H}-H\frac{\ddot{\chi}^{(0)}}{\dot{\chi}^{(0)}}\right)\,.
\label{barPsi_barPsi2}
\ee

We have now almost all the ingredients needed to evaluate the different terms involved in 
the backreaction equation (\ref{EQ1simpl_2}). Let us begin with 
$\langle \bar{\psi} \dot{\bar{\psi}}\rangle$:
\be
\langle \bar{\psi} \dot{\bar{\psi}}\rangle=\frac{H^2}{\dot{\chi}^{(0)\,2}}\left[
\left(\frac{\dot{H}}{H}-\frac{\ddot{\chi}^{(0)}}{\dot{\chi}^{(0)}}\right)\langle \chi^{(1)\,2}
\rangle+\langle \chi^{(1)} \dot{\chi}^{(1)} \rangle\right]\,.
\label{psi_psidot}
\ee

Following \cite{FMSVV2} we solve the background equations for our particular model 
and find the following zero order solution for the test field $\chi$
\be
\chi^{(0)}(t)=\chi^{(0)}(t_i)\left(
\frac{H(t)}{H(t_i)}\right)^{\alpha} \,, 
\label{test_field}
\ee
where we have defined $\alpha=\frac{m_\chi^2}{m^2}$.

Let us remark that, for the purpose of this work, based on a perturbative analysis,
we need to consider $\chi^{(0)}(t_i)\ne 0$, namely, a dynamical regime such that
$\chi$ has non zero vacuum expectation value, a condensate.
In such a case we have a scalar with time-like gradient which defines a space-like
hypersurface, $\chi(x)$ equal to a constant, up to when the space dependence of the scalar
can be seen as a perturbative dependence over a time function $\chi^{(0)}(t)$.
Namely, when the stochastic perturbative approach defined in \cite{FMSVV2} is valid (see, Eqs.
(\ref{condphi}-\ref{condchi})). Such a configuration is compatible with a dynamical phase
of universe expansion such as inflation.
Let us further underline as the dynamical cut-off, present in the stochastic approach of \cite{FMSVV2} and
which is used to regularize our correlator (as, for example, $\langle \chi^{(1) 2}\rangle$), is defined in the way
that only the super-Hubble ''classical'' fluctuations of our fields are taken in consideration.
If the background value of $\chi$ were to vanish, $\chi$ would not define a physical clock field
which can be used to follow the evolution of space-time.

At this point we turn to the evaluation of the contribution of the long wavelength
(super-Hubble) modes to the expectation values which appear in Eq. (\ref{psi_psidot}).
Following \cite{FMSVV2} we obtain for our particular background the following 
stochastic solutions up to second order in perturbation theory
\begin{eqnarray}
\langle \chi^{(1) 2}\rangle &=& \frac{3 H^{2 \alpha}}{8
\pi^2 m^2 (2-\alpha)} ( H_i^{4-2 \alpha} \!- \!H^{4-2 \alpha} )
-\frac{\alpha^2}{48 \pi^2}\frac{\chi^{(0)}(t_i)^2}{M_{pl}^2}
\left(\frac{H}{H_i}\right)^{2 \alpha}\frac{1}{H^4}\left(H^2-H_i^2\right)^3\,,
\label{CorrChiPart}
\end{eqnarray}

\begin{eqnarray}
\langle \chi^{(2)} \rangle &=&\frac{\alpha}{8 \pi^2}
\frac{\chi^{(0)}(t_i)}{M_{pl}^2} \left(\frac{H}{H_i}\right)^{\alpha}
\left[-\frac{H_i^6}{H^4}\frac{1-\alpha/2}{6}+\frac{H_i^4}{H^2}
\frac{1-\alpha}{4} 
+H_i^2 \frac{\alpha}{4}
-H^2\frac{1+\alpha}{12}\right] \,.
\label{Chi2ord}
\end{eqnarray}

Going back to the expression of Eq.(\ref{psi_psidot}) we obtain by substitution 
the following result
\begin{eqnarray}
\langle \bar{\psi} \dot{\bar{\psi}}\rangle&=& \frac{H^2}{\dot{\chi}^{(0)\,2}}\frac{\dot{H}}{H}
\left\{\frac{3 H^{2 \alpha}}{8
\pi^2 m^2 (2-\alpha)} \left[\left(2-H\frac{\ddot{H}}{\dot{H}^2}\right)H_i^{4-2 \alpha}
-\left(4-\alpha- H\frac{\ddot{H}}{\dot{H}^2}\right)H^{4-2 \alpha}\right]\right.\nonumber \\
& & \left. -\frac{\alpha^2}{48 \pi^2}\frac{\chi^{(0)}(t_i)^2}{M_{pl}^2}
\left(\frac{H}{H_i}\right)^{2 \alpha}\frac{1}{H^4} \left[-H\frac{\ddot{H}}{\dot{H}^2}\left(H^2-H_i^2\right)^3+3 H^2 \left(H^2-H_i^2\right)^2\right]
\right\}\,.
\label{PsiPsiDot}
\end{eqnarray}
 
Let us now consider the other term in Eq.~\eqref{EQ1simpl_2} which depends on 
$\langle \dot{\bar{\psi}}^{(2)}\rangle$. 
It is easy to show that the leading slow-roll contribution is given by the derivative of the 
leading term of $\langle {\bar{\psi}}^{(2)}\rangle$. 
Indeed, considering the leading part in Eq.(\ref{barPsi_barPsi2}), then in the slow-roll 
approximation, and starting from the expression
\be
\langle {\bar{\psi}}^{(2)}\rangle\simeq\frac{H}{\dot{\chi}^{(0)}} \langle\chi^{(2)}\rangle-
\frac{H^2}{\dot{\chi}^{(0)\,2}}\langle\chi^{(1)\,2}\rangle\,,
\ee
one obtains that
\be
\langle \dot{\bar{\psi}}^{(2)}\rangle\simeq
\frac{H}{\dot{\chi}^{(0)}} \left[\left(\frac{\dot{H}}{H}-\frac{\ddot{\chi}^{(0)}}{\dot{\chi}^{(0)}}\right)\langle \chi^{(2)}
\rangle+\langle \dot{\chi}^{(2)} \rangle\right]-2 \langle \bar{\psi} \dot{\bar{\psi}}\rangle\,,
\ee
from which it follows that the leading value of $\langle \dot{\bar{\psi}}^{(2)}\rangle$
will be given by 
\be
\langle \dot{\bar{\psi}}^{(2)}\rangle\simeq
\frac{H}{\dot{\chi}^{(0)}} \left[-2\frac{\dot{H}}{H}-
\frac{\ddot{H}}{\dot{H}}\right]\langle \chi^{(2)}
\rangle-2 \langle \bar{\psi} \dot{\bar{\psi}}\rangle\,.
\ee
As a consequence, we can rewrite the cosmological backreaction
(Eq.(\ref{EQ1simpl_2})) for the expansion rate as seen by an observer comoving 
with the light test field $\chi$ in the following way
\be
H_{eff}^2\simeq
H^2 \left[1+\frac{6}{H}\langle \bar{\psi}\dot{\bar{\psi}}\rangle+
\frac{2}{\dot{\chi}^{(0)}}\left(2\frac{\dot{H}}{H}+\frac{\ddot{H}}{\dot{H}}\right)
\langle \chi^{(2)}\rangle\right]\,.
\label{BR_H_2_field_a}
\ee
This equation will be the starting point of the analysis which we shall develop in Section $4$.

\subsection{Bounds}

Let us now see which are the constraints that we have to take into consideration 
to have a consistent description of our problem.

{\it Test field condition for $\chi$.}
The condition that the gravitational background dynamics is not affected by the field $\chi$
can be derived from $\rho_\chi\ll \rho_\Phi$ using the solution in Eq.~\eqref{test_field} and, 
for the whole inflationary period, reads~\cite{FMSVV2}
\be 
\chi^{(0)}(t_i)^2 \ll \left[1+\frac{\alpha}{9}\frac{m^2}{H^2}\right]^{-1}
\frac{1}{\alpha}\left(\frac{H}{H_i}\right)^{2-2\alpha}6
\frac{H_i^2}{m^2} M_{pl}^2 \,.
\label{test_field_condition}
\ee 
In particular, for the case $\alpha \ll 1$, we
obtain the following limiting condition at the end of inflation
($H \simeq m$):
\be 
\chi^{(0)}(t_i)^2 \ll \frac{6}{\alpha}
M_{pl}^2 \,.
\label{ModuliCond} 
\ee 

{\it Reliability of the perturbative approach.}
We study the quantum fluctuations of the system of fields $\Phi$ and $\chi$ in perturbation 
theory. Such an approach is valid provided that the perturbed quantities are smaller in 
amplitude than the background values. There are
several conditions which need to be satisfied
\bea
&{}&\frac{\left(\langle \phi^{(1) 2}\rangle\right)^{1/2}}{\phi^{(0)}}\ll 1 
\,\,\,\,\,\,\,\,,\,\,\,\,\,\,\,\,
\frac{\langle \phi^{(2)} \rangle}{\left(\langle \phi^{(1) 2}\rangle\right)^{1/2}}\ll 1 \,,\label{condphi}\\
&{}&\hspace{2cm}\frac{\langle \phi^{(1)} \chi^{(1)}\rangle}{ \phi^{(0)}\chi^{(0)} }\ll 1\,,\label{condmix}\\
&{}&\frac{\left(\langle \chi^{(1) 2}\rangle\right)^{1/2}}{\chi^{(0)}}\ll 1 
\,\,\,\,\,\,\,\,,\,\,\,\,\,\,\,\,
\frac{\langle \chi^{(2)} \rangle}{\left(\langle \chi^{(1) 2}\rangle\right)^{1/2}}\ll 1 \,.
\label{condchi}
\eea
The first conditions given in~\eqref{condphi} are related only to the inflaton field. For the 
chaotic model under investigation they have been already analyzed in the appendix of 
\cite{FMSVV1}. One finds that
the perturbative treatment up to second order of the inflaton field is reliable, for all the 
duration of inflation, only for a value of
$H_i$ less of nearly $130 m$ (where we consider $M_{pl}=10^5 m$).

It can be shown that the last two conditions (Eq.~\eqref{condchi}), which involve only the 
test field $\chi$, are the most stringent, and in particular stronger that the one which 
contains the mixed correlator in Eq.~\eqref{condmix}.

In Section $4$ we shall present a numerical evaluation based only on the constraints in 
(\ref{condchi}), together with the one in Eq.(\ref{test_field_condition}).
We shall also introduce a bound to control the magnitude of the backreaction (see Eq.~\eqref{BR_H_2_field_a}). Indeed, to obtain a reliable estimate of the quantum backreaction 
in a perturvative way, it is natural to study a contraint to limit its value. If the backreaction
terms become comparable to the background values, then the perturbative approach
will break down.

\subsection{Estimates for the backreaction}

Before presenting the numerical evaluation of the backreaction effect it is useful to 
investigate the analytical expressions of the various contributions to the effective
Hubble expansion rate in order to extract the leading behaviour in the different phases 
of the inflationary evolution, in particular during the final stages of inflation. Since
the phase space of infrared modes is increasing as we approach the end of inflation,
we expect the magnitude of the backreaction terms to increase.

Let us begin with the first term of the cosmological quantum backreaction 
terms in Eq.~\eqref{BR_H_2_field_a}, the one proportional to 
$\langle \bar{\psi} \dot{\bar{\psi}}\rangle$.
At the end of inflation, the leading value of the second term in Eq.~\eqref{PsiPsiDot}  is
negligible with respect to the leading value of the first one, for
$\alpha<2$ which is the region of interest for our model since we want to describe light fields, 
provided that
\be
\chi^{(0)}(t_i)^2 \ll \frac{81}{\alpha^2} \frac{2}{2-\alpha} \frac{M_{pl}^2 m^2}{H_i^2}\,.
\label{Condition2}
\ee
Note that this condition is different from the condition (\ref{ModuliCond}). 
If we consider the particular case $\alpha\ll 1$ and require that
(\ref{ModuliCond}) implies (\ref{Condition2}), we obtain the
following condition on $\alpha$: 
\be 
\alpha < \frac{27}{2}
\frac{m^2}{H_i^2}\,. 
\label{Limitonalpha}
\ee
In such a case the leading result is given by
\be
\frac{6}{H} \langle \bar{\psi} \dot{\bar{\psi}}\rangle = \frac{\dot{H}}{\dot{\chi}^{(0)\,2}}
\frac{9 H^{2 \alpha}}{2
\pi^2 m^2 (2-\alpha)}  H_i^{4-2 \alpha} \, .
\label{LeadingPsiPsiDot}
\ee
Note that the sign of this contribution is negative. Hence, this backreaction term
will contribute to a decrease in the measured local expansion rate, an effect
conjectured in \cite{ABM, RHBrev}. The contribution is increasing as inflation proceeds, an effect which is
due to the increasing phase space of infrared modes, 
before being attenuated at the end of inflation by an $H^2$ factor (coming from the $1/(\dot{\chi}^{(0)})^2$). 
Returning to the full expression
of Eq.~\eqref{PsiPsiDot}, let us note that at the beginning of inflation the second
term in the square parentheses of the first line dominates over the first, and that
therefore the sign is opposite, and thus the contribution to the effective Hubble
expansion rate starts out positive.

We are left with the task of  analyzing  the magnitude of the  second term in 
Eq.~\eqref{BR_H_2_field_a}, proportional to $\langle \chi^{(2)}\rangle$.
Starting from Eq.(\ref{Chi2ord}), one can see that this term reaches
its maximum value at the end of inflation or more generally for $H\ll H_i$. 
Working in this limit, assuming $\alpha\ll2$, considering the long wavelength limit
and calculating in leading slow-roll approximation, one finds
\be
\frac{2}{\dot{\chi}^{(0)}}\left(2\frac{\dot{H}}{H}+\frac{\ddot{H}}{\dot{H}}\right)
\langle \chi^{(2)}\rangle\simeq - \frac{1}{12 \pi^2} \left(1-\frac{\alpha}{2}\right) 
\frac{H_i^6}{M_{pl}^2 H^4}\,.
\label{Contr_2}
\ee
A first interesting observation is that this contribution (\ref{Contr_2}) is always 
negative  (as well as the full expression of $\langle \chi^{(2)}\rangle$ in Eq.~\eqref{Chi2ord})
and once again goes in the direction of decreasing the effective expansion rate of the 
Universe given in Eq.~\eqref{BR_H_2_field_a}, 
namely the expansion rate measured by the observer "sitting" on $\chi$. 
Furthermore, this contribution
is of the same order of magnitude of the one found in \cite{MV} for the case of a single 
field model. In fact, in \cite{MV} we found, at  leading order in the slow-roll 
approximation and in the long wavelength limit, a backreaction effect
on the expansion rate of the Universe as seen by an isotropic observer, of the 
following magnitude:
\be
\frac{\dot{H}}{H^2} \frac{\langle \phi^{(1) 2} \rangle}{M_{pl}^2} \simeq -\frac{1}{24 \pi^2} \frac{H_i^6}{M_{pl}^2 H^4}
\label{Contrc=1}
\ee
which is indeed of the same order as (\ref{Contr_2}). 

We have now seen that both backreaction terms in Eq.~\eqref{BR_H_2_field_a} reduce the
effective expansion rate of the Universe as seen by an observer located on constant $\chi$
surfaces. We now turn to an estimate of the magnitude of the backreaction effect
in the region of parameter space in which the bounds presented in the previous subsection
apply. We consider only the final stage of the inflationary era ($H\ll H_i$) and, for simplicity, 
maintain the condition of Eq.(\ref{Condition2}).

Using Eq.(\ref{test_field}) and the slow-roll approximation $\dot{H}\simeq- m^2/3$, 
the expression in Eq.~\eqref{LeadingPsiPsiDot} becomes
\be 
\frac{6}{H} \langle \bar{\psi}\dot{\bar{\psi}}\rangle \simeq 
-\frac{1}{{\chi}^{(0)}(t_i)^{2}} \frac{27}{2 \pi^2} \frac{1}{\alpha^2 (2-\alpha)}
\frac{H^2 H_i^4}{m^4}\,.
\label{Contr_1}
\ee
If we now consider, in order to be consistent, the test field condition (\ref{ModuliCond}) 
(considering the limit (\ref{Limitonalpha}) in the case when this implies the condition (\ref{Condition2})) 
we find a lower bound to the above contribution 
\be  
\left| \frac{6}{H} \langle \bar{\psi}\dot{\bar{\psi}}\rangle \right|\gg
\frac{9}{4 \pi^2}\frac{1}{\alpha (2-\alpha)}\frac{H^2 H_i^4}{M_{pl}^2 m^4}\,.
\label{Contr_1_est}
\ee
This is to be compared with the other contribution to the backreaction explicity given in 
Eq.(\ref{Contr_2}) for the case $H \ll H_i$. 
 
Concerning their order of magnitude we note that, under the condition considered here, 
while  the contribution (\ref{Contr_1}) depends,
for the given model, on the zeroth order value ${\chi}^{(0)}_i$ and on the mass 
$m_\chi$ of the test field, the contribution (\ref{Contr_2}) is almost independent of the 
features of the test field, provided that it is light ($\alpha\ll1$).
Furthermore, for typical values ​​of our parameters we will have 
\be 
\left| \frac{6}{H} \langle \bar{\psi}\dot{\bar{\psi}}\rangle\right| \gg \left|
\frac{2}{\dot{\chi}^{(0)}(t_i)}\left(2\frac{\dot{H}}{H}+\frac{\ddot{H}}{\dot{H}}\right)
\langle \chi^{(2)}\rangle\right|
\ee
and the first contribution will be the leading one.
For example, if we consider a light test field at the end of inflation, and the following 
typical value for a chaotic model of inflation $H_i={\cal O} (10) m$ with 
$m=10^{-5} M_{pl}$, we obtain (at the end of inflation when $H \sim m$) the 
following order of magnitude (from Eqs. (\ref{Contr_1_est}) and (\ref{Contr_2}))
\be  
\left| \frac{6}{H} \langle \bar{\psi}\dot{\bar{\psi}}\rangle \right|\gg
\frac{1}{\alpha (2-\alpha)} {\cal O} (10^{-6})
\ee
\be
\left|\frac{2}{\dot{\chi}^{(0)}(t_i)}\left(2\frac{\dot{H}}{H}+\frac{\ddot{H}}{\dot{H}}\right)
\langle \chi^{(2)}\rangle\right|\simeq {\cal O} (10^{-6})\,.
\ee

\begin{figure}[h]
\centering
\includegraphics[height=.25\textwidth]{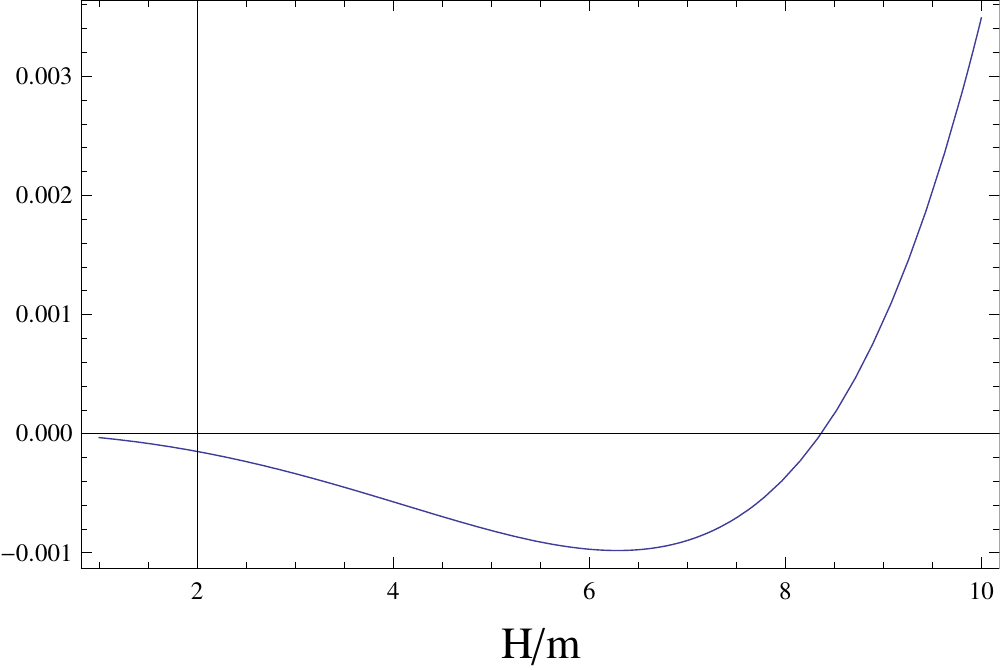} \hspace{13mm} \includegraphics[height=.25\textwidth]{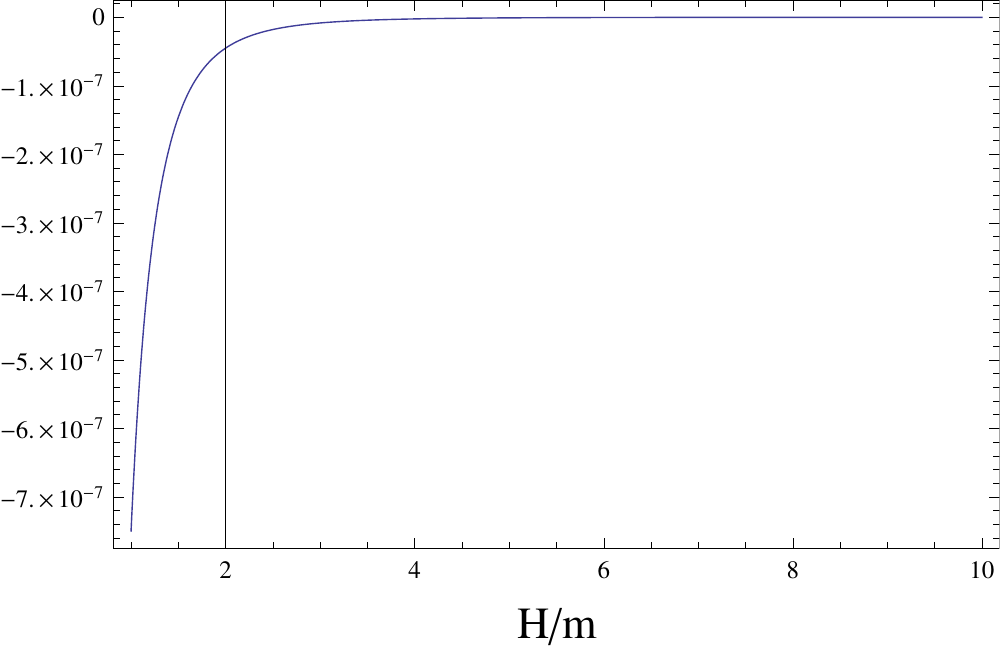}
\caption{We show $\frac{6}{H}\langle \bar{\psi}\dot{\bar{\psi}}\rangle$ (left plot) and $\frac{2}{\dot{\chi}^{(0)}}\left(2\frac{\dot{H}}{H}+\frac{\ddot{H}}{\dot{H}}\right)
\langle \chi^{(2)}\rangle$ (right plot) as functions of $H/m$ with $H_i=10 m$, $\alpha=0.2$ and $\chi_0=0.7 M_{pl}$.
}
\label{fig1}
\end{figure}

During the early stages of inflation the behaviour of 
$\frac{6}{H} \langle \bar{\psi}\dot{\bar{\psi}}\rangle$ is more complicated, 
but in general it will be larger in amplitude than 
$\frac{2}{\dot{\chi}^{(0)}}\left(2\frac{\dot{H}}{H}+\frac{\ddot{H}}{\dot{H}}\right) \langle \chi^{(2)}\rangle$. 
In particular, as said, looking at the dominant contribution in the first line of \eqref{PsiPsiDot}, 
one can see that it is positive when the second term in the square bracket overcomes the 
first one. This happens approximately up to the point where $H\simeq 2^{-1/(4-2\alpha)} H_i$. 
The ratio between the maximum positive backreaction contribution to $H_{eff}^2$ and the 
absolute value of the maximum negative contributions is monotonically decreasing from 
approximately $3.7$ ($\alpha\ll 1$) to $3.2$ ($\alpha=1$). 
In Fig.~\ref{fig1}  we plot these two different contributions, to the
effective Hubble expansion rate, without making any approximation, and for typical 
values of our parameter for which the backreaction is still perturbative and our results 
consistent.

Furthermore, for a very light test field, 
the contribution of the first term $\sim  \frac{6}{H} \langle \bar{\psi}\dot{\bar{\psi}}\rangle$ may result  
(since it contains a $1/\dot{\chi}^{(0)\,2}$ enhancement  factor (see Eq.(\ref{PsiPsiDot})))
in a value which is even greater in magnitude than $1$. This may happen even if the 
perturbative expansion used to study the dynamics is still valid. But the perturbative
approach to study backreaction will have broken down.
We shall make more comment on this fact in Section $4$.

\subsection{Gauge invariant backreaction on the effective equation of state}

A valuable piece of information, which can be obtained without much effort, 
is given by the effective equation of state with respect to observers which
measure the effective Hubble expansion rate discussed here.
For this purpose, we take advantage of the fact that one of the quantum 
contributions (the second one) to the effective expansion rate in
Eq.~(\ref{BR_H_2_field_a}) can be written as $c \,(\dot{H}/H^2)\, \langle \varphi^2 \rangle/M_{pl}^2$ 
with $c=2-\alpha$ (see Eqs.(\ref{Contr_2}, \ref{Contrc=1})).
Let us start from the general result obtained in \cite{MV}, valid for any class 
of observers and slow-roll inflationary models, 
where we allow for a further second order contribution $H^2 B$ in the 
effective expansion rate, beyond the typical terms $\sim \langle \varphi^2 \rangle$,
\be
\dot{\chi}^{(0)\,2} \left(\frac{1}{a_{\eff}}\frac{\partial \, a_{\eff}}{\partial \chi_0}
\right)^2 \equiv \frac{8\pi G}{3} \rho_{eff}=
H^2\! \left[1\!+ \left( 
c \frac{\dot{H}}{H^2}+ d \frac{\dot{H}^2}{H^4} +
{\cal O} \left(\frac{\dot{H}^3}{H^6}\right)\right)
\frac{\langle \varphi^2 \rangle}{M_{pl}^2}+B(t)\right]\,.
\ee
In the single field model~\cite{MV}, the parameters $c$ and $d$ encode the 
typical non zero backreaction effects at first and second order in the slow-roll approximation  
and $B(t)$ encodes other possible effects (in our case $c=2-\alpha$ while $d$ is not fixed). 
Since this latter quantity is of  second order in perturbation theory we have to consider only 
linear terms in any perturbative expression.
Then, from the consistency between the effective equations for the averaged geometry 
(see \cite{FMVVprl,MV}), one obtains
\begin{eqnarray}
-\frac{1}{a_{\eff}}\dot{\chi}^{(0)} \frac{\partial }{\partial \chi_0}\left(\dot{\chi}^{(0)} \frac{\partial }{\partial \chi_0} \, a_{\eff} \right)\!\!\!\! &\equiv& \!\!\!\!  \frac{4\pi G}{3} \left(\rho_{eff}+3 p_{eff} \right)=
-\dot{H}\!-\!H^2\!-\!H^2 \!\left\{\left[c \frac{\dot{H}}{H^2} +
\left(d-\frac{3}{2} c\right) \frac{\dot{H}^2}{H^4}
\right. \right. \nonumber \\
& & \left. \left.\!\!\!\!
+ c \frac{\ddot{H}}{H^3}+{\cal O} \left(\frac{\dot{H}^3}{H^6}\right)\right]
\frac{\langle \varphi^2 \rangle}{M_{pl}^2} +B(t)+\frac{\dot{H}}{2 H^2} \left[B(t)+\dot{B}(t)\right]\right\}\,.
\end{eqnarray}
From these relations it is easy to see that the effective equation of state $w_{\eff}=p_{\eff}/\rho_{\eff}$,
to first non trivial order, is given by
\be
w_{\eff}=\frac{p_{\eff}}{\rho_{\eff}}=-1-\frac{2}{3} \frac{\dot{H}}{H^2}
+\left\{\left[\frac{5}{3} c \frac{\dot{H}^2}{H^4}-\frac{2}{3} c \frac{\ddot{H}}{H^3}+{\cal O} \left(\frac{\dot{H}^3}{H^6} \right)\right]
\frac{\langle \varphi^2 \rangle}{M_{pl}^2}+\frac{\dot{H}}{3 H^2} B(t)-\frac{1}{3 H} \dot{B}(t)\right\}\,,
\label{ESeffective}
\ee
where the $d$ dependence disappears in the leading order correction.
In the case under investigation
\be 
B=\frac{6}{H} \langle \bar{\psi}\dot{\bar{\psi}}\rangle \, .
\ee

\section{Discussions and numerical results}

In the previous section we have defined our model, and studied the main observables 
in a GI way, giving their analytical expressions for the quantum corrections. Having 
also provided the general criteria for the reliability of our model within a perturbative 
approach, we want now to analyze numerically the magnitude of the overall backreaction 
effects experienced by our observers due to quantum fluctuations.

Let us first begin with some general considerations.
As seen, in the evaluation of the backreaction from Eq.(\ref{EQ1simpl_2}), the 
quantities $\bar{\psi}$ and $\bar{\psi}^{(2)}$ are connected to $\chi^{(1)}$ and 
$\chi^{(2)}$ by the gauge trasformation given in Eqs.(\ref{GT1},\ref{GT2},\ref{GT3}). 
Therefore, the backreaction terms will be given by combinations of test field perturbations 
multiplied by factors like $1/\dot{\chi}^{(0)}$ and $(1/\dot{\chi}^{(0)})^2$.
These factors replace the similar factors $1/\dot{\phi}$ and  $(1/\dot{\phi})^2$
which are present in the backreaction terms obtained in \cite{MV} for the single field model.
On the other hand, even in comparison with the slow rolling of the background 
inflaton field, the background test field $\chi^{(0)}$ changes relatively slowly for 
$\alpha \ll 1$ (see Eq.(\ref{test_field})) and hence the above factors 
can give a strong enhancement of the backreaction compared to the effect of the factors $(1/\dot{\phi})^2\sim 1/M_{pl}^2$ in the case of single field inflation. 
So, contrary to \cite{FMSVV2}, where it was shown that for a chaotic model 
the mean square gauge-invariant inflaton fluctuations grow faster than  those of any
test field with nonnegative mass, 
the backreaction on the effective expansion rate, measured with respect to the 
observer comoving with the light test field in a two field model, is at least of the same order, and
in general bigger, than the typical backreaction that we can obtain in a single field case.

The relative magnitude of the backreaction terms in the expression for the effective
Hubble expansion rate has already been given in Fig. 1. We now want to numerically
study the parameter space where our analysis is under perturbative control.

Concerning the validity of the expressions for the quantum corrections (backreaction) 
found in the previous section, we have a safe perturbative expansion for the observables 
only for $\frac{6}{H} \langle \bar{\psi}\dot{\bar{\psi}}\rangle\ll 1$, since this appears 
in the leading terms both in Eq.(\ref{BR_H_2_field_a}) and in Eq.(\ref{ESeffective}).

In the following we give the limit of validity of our results imposing the condition that the
backreaction contributes a fixed maximal amount to the effective Hubble expansion rate,
i.e. $|H_{eff}^2-H^2|/H^2\ll 1$. In the numerical analysis we consider the
condition $|H_{eff}^2-H^2|/H^2<10^{-2}$. To do this we join the test field condition 
(\ref{test_field_condition}) (with the $\ll$ sign replaced by $<10^{-2}$) and conditions
on the validity of the perturbative expansion of the fields. The latter can be obtained by 
requiring the condition given in Eq.~\eqref{condchi} (with the $\ll$ sign replaced by $<10^{-1}$), 
which is the most stringent one among~\eqref{condphi}-\eqref{condchi}.

\begin{figure}[ht!]
\centering
\includegraphics[height=.35\textwidth]{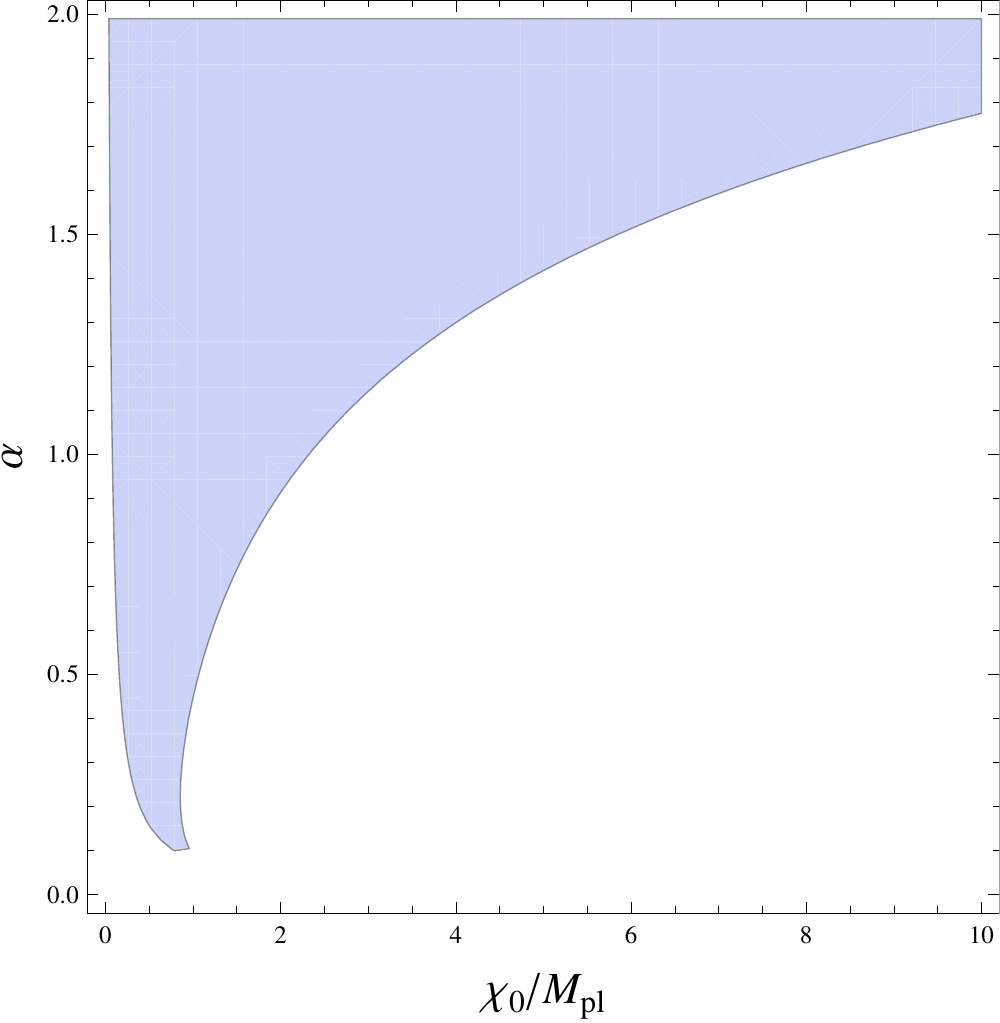}\hspace{2cm}\includegraphics[height=.35\textwidth]{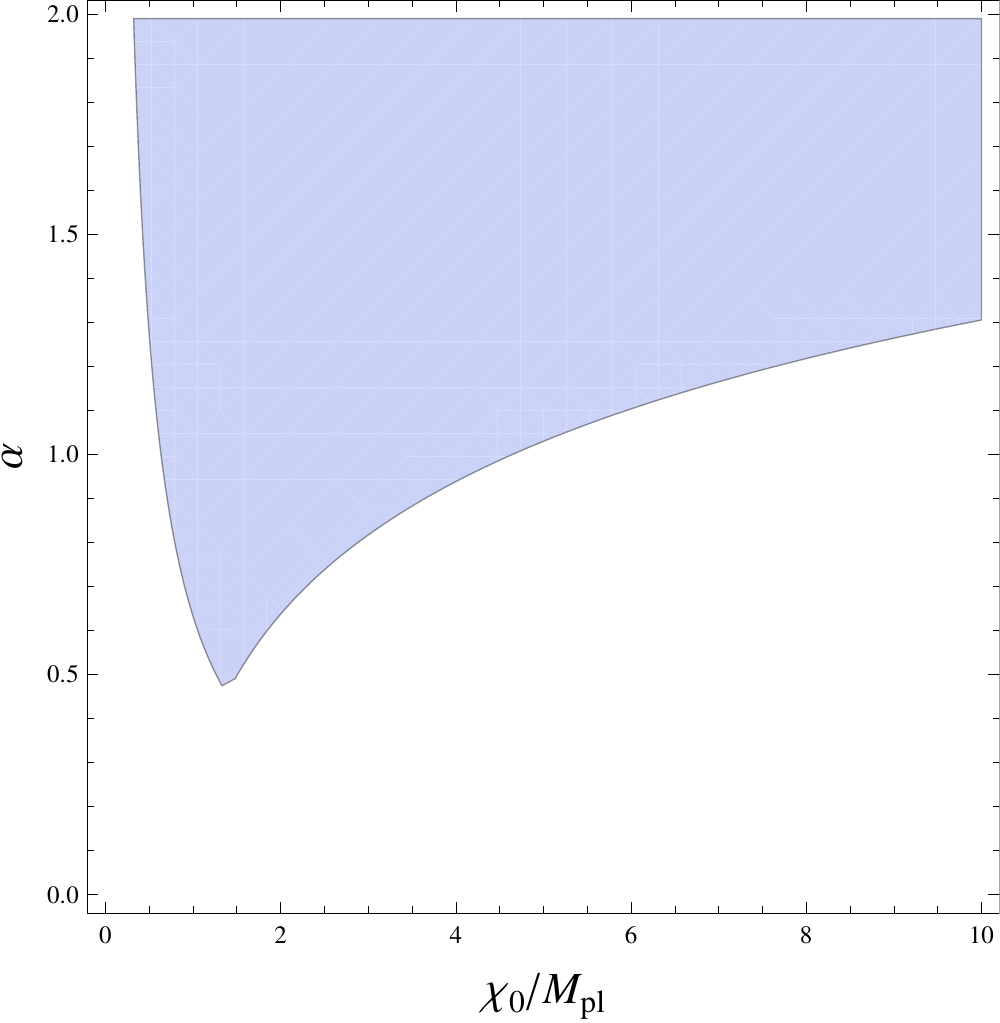}
\caption{We plot the region where all three conditions (validity of the perturbation theory, test 
field approximation, both defined in the text, and requiring the backreaction to be at most 
$1\%$) are satisfied, for the values $H_i=10 m$ (left plot) and $H_i=20 m$ (right plot). The
horizontal axis denotes ${\chi}^{(0)}(t_i)/M_{pl}$ and the vertical axis $\alpha$. We
chose the value of $m$ given by $M_{pl}=10^5 m$ to be consistent with the observed value of
cosmological perturbations.
\label{fig2}
}
\end{figure}

\begin{figure}[ht!]
\centering
\includegraphics[height=.35\textwidth]{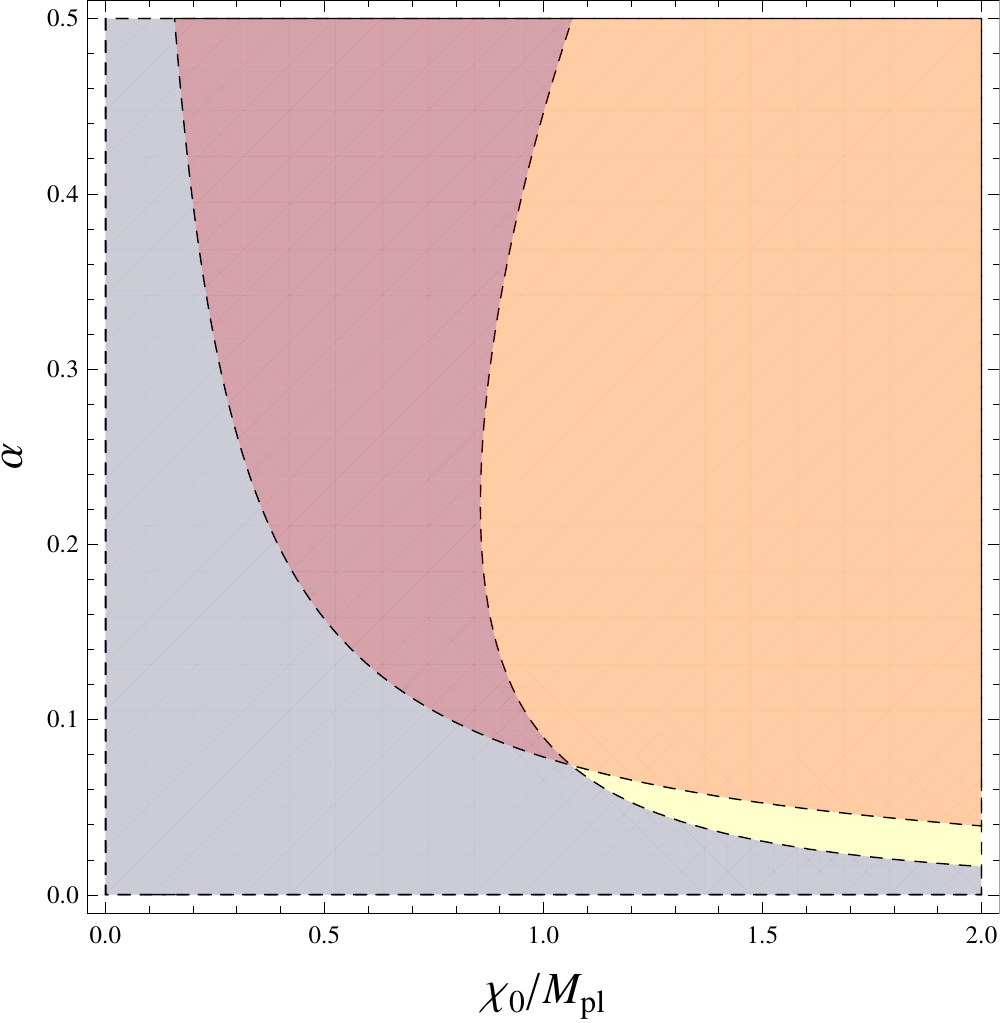}
\hspace{2cm}
\includegraphics[height=.35\textwidth]{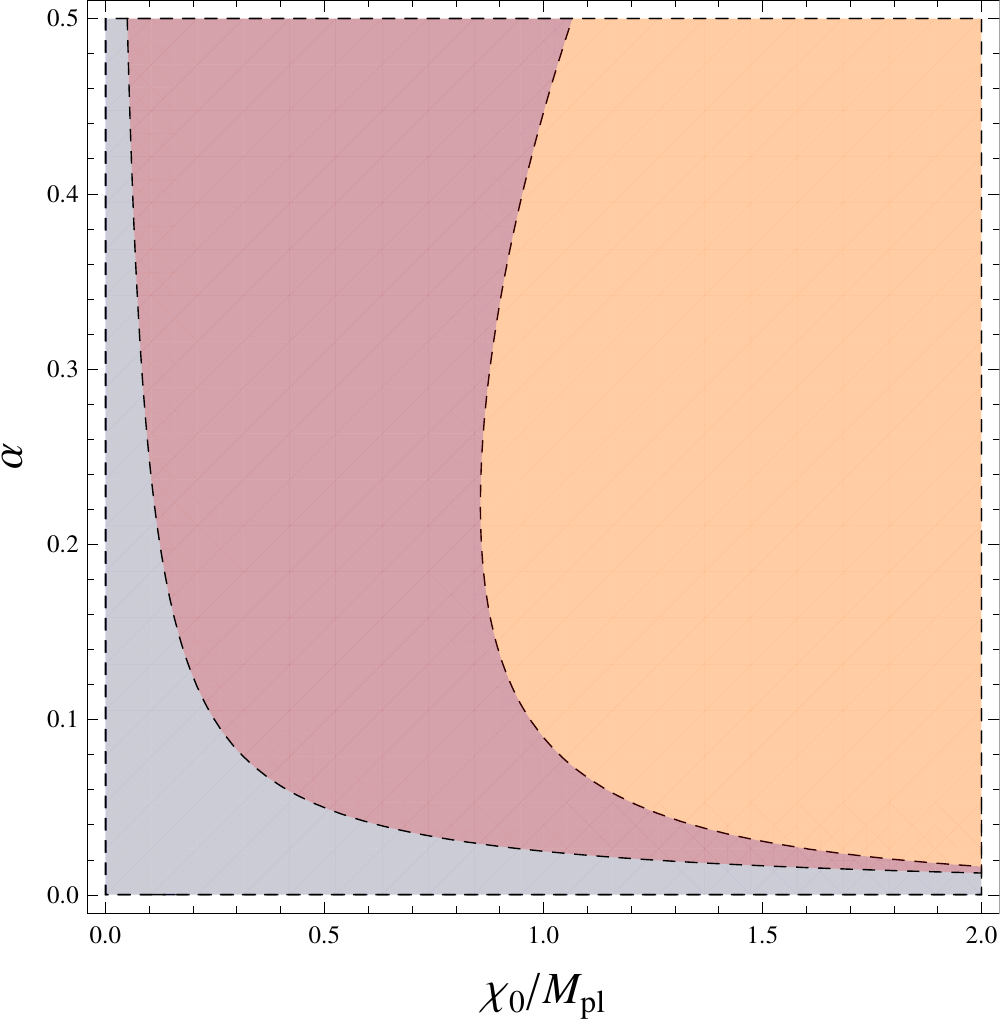}
\caption{The darker region in each plot corresponds to the intersection of
 the regions where the three conditions (validity of the perturbation theory, test field approximation, 
upper bound of the backreaction at $1\%$ (left plot) or $10\%$ (right plot)) are satisfied. We 
have $H_i=10 m$ and label the plane with ${\chi}^{(0)}(t_i)/M_{pl}$ on the  horizontal axis and 
$\alpha$ on the  vertical axis ($M_{pl}=10^5 m$).}
 \label{fig3}
\end{figure}

In Fig.~\ref{fig2} we consider the cases of $H_i=10 m$ and $H_i=20 m$, and show the
region in the two-dimensional parameter space labelled by $\alpha$ and
by the initial value of the $\chi$ field for which all of the conditions are
satisfied. The shaded region is the allowed one. For very small initial values of
the $\chi$ field the backreaction becomes too large in amplitude, for very large
values the test field condition is satisfied only for $\alpha \ll 1$ (see also Fig \ref{fig3}). 
This leaves us with the indicated
region. For higher values of $H_i$ the region narrows more and more until it disappears 
around a value of $H_i\simeq 50 m$.

We have still not yet discussed the change in magnitude of the backreaction as our
parameters vary. In the region where the backreaction is small, then for the whole duration of 
inflation we can study numerically how it depends on the free parameters of the 
model (initial condition $\chi^{(0)}(t_i)$ and the mass $m_\chi$).
We find it convenient to show in Fig.~\ref{fig3} how the constrained region in parameter space, 
obtained requiring a certain amount of maximum backreaction, changes as the limit on
the amplitude of backreaction is changed. In each of the plots, the right dashed line separates
the region where the test field approximation is valid (left of the line) from the region where it
is not (right of the line). The left dashed line separates the region where the backreaction
effect on the observed Hubble parameter is smaller than the limit indicated (right of the line)
from that where it is not (left of the line). One can see that the relative amplitude of
backreaction on the local Hubble expansion rate increases for smaller values of
the mass $m_\chi$ and for smaller $\chi^{(0)}(t_i)$.

In both the two plots of Fig.\ref{fig3} we give also further informations related to the 
different bounds. Each bound is associated to a region which has a colour: yellow 
(perturbation theory valid), blue (test field approximation) and light red (backreaction 
with an upper bound). Where two regions overlap the colours change as follow: blue+yellow 
is gray, yellow + light red is orange. Where all the bounds are satisfied, all regions overlap, 
the colour is darker red.

It may be interesting to follow the results obtained in our perturbative analysis up to the point 
when the backreaction effects are of the same order (positive or negative) of the background, 
while maintaining the other bounds (test field approximation and validity of the cosmological 
perturbation theory for metric and matter fluctuations).
We want to stress that this is of course not formally a safe choice since the backreaction on 
the observables might have important perturbative contributions at higher order. However, it
may be useful to gain some qualitative indication of the effects.

We have therefore studied, for several values of the parameters $\alpha$ and $\chi^{(0)}(t_i)$, 
$H_{eff}$ and $w_{eff}$ defined by the observer comoving with the light test field. It is 
important to stress that even assuming that such observations might be have a physical interest
during inflation, they are not connected in a simple way to late time observations, unlike
the direct relation which exists in ordinary cosmological perturbation theory for the spectrum 
of the cosmological fluctuations.

\begin{figure}[ht!]
\centering
\includegraphics[height=.25\textwidth]{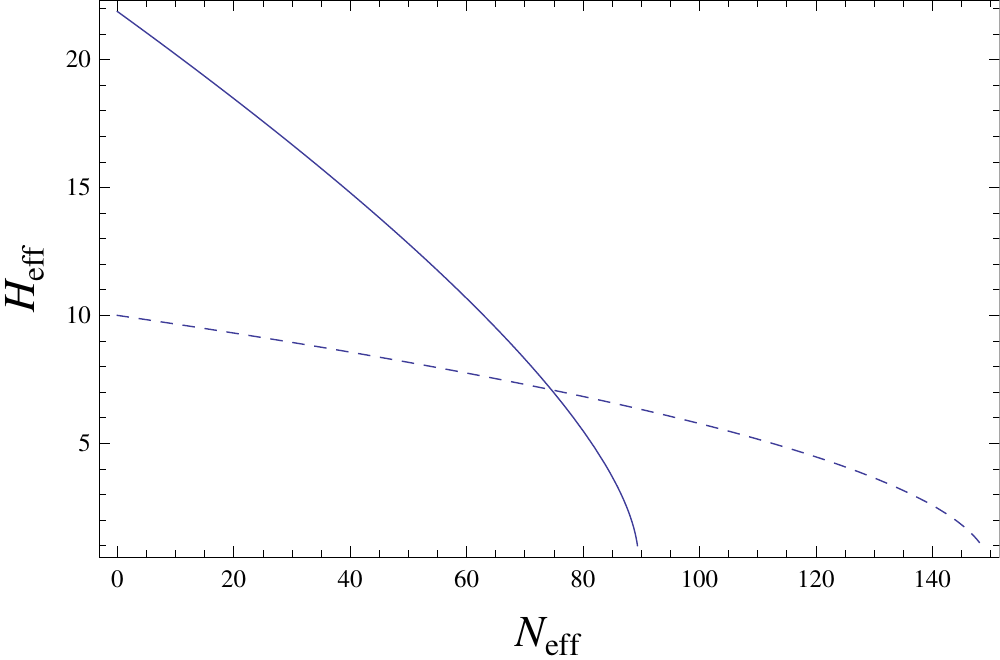} \hspace{1.5cm} \includegraphics[height=.25\textwidth]{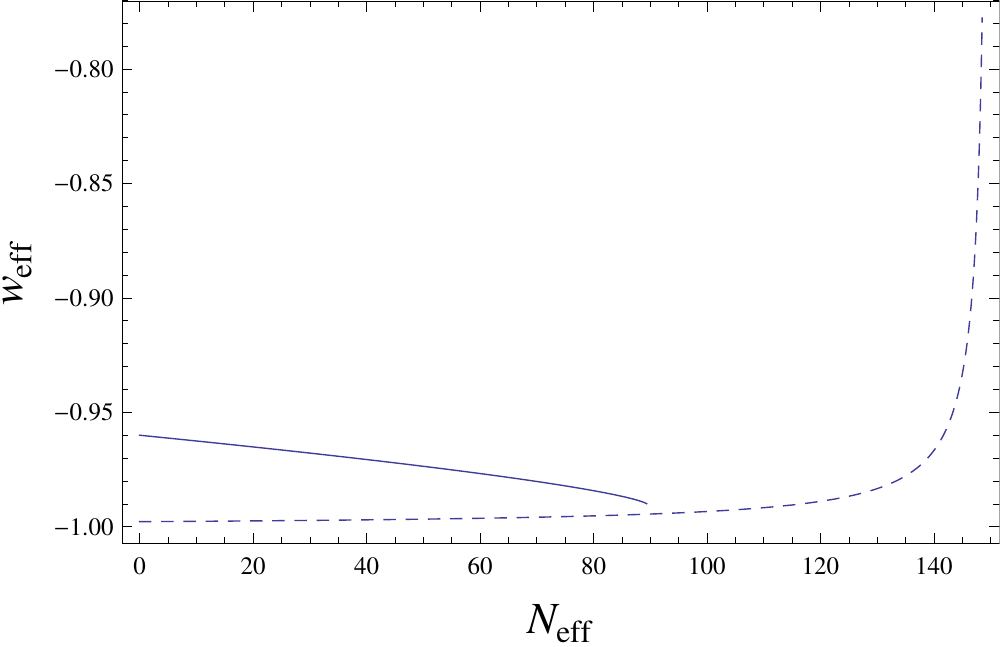}
\caption{The effective expansion rate (left plot) and the equation of state parameter $w_{eff}$
(right plot), which both include the quantum backreaction, measured by an observer comoving 
with the test field $\chi$, as a function of the associated effective number of e-folds of
inflation (continuous lines) are shown together with the corresponding background 
values versus the standard number of e-folds (dashed lines). We have chosen 
$H_i=10 m$, $\alpha=0.0085$ and $\chi_0=0.5 M_{pl}$ ($M_{pl}=10^5 m$).}
 \label{fig4}
\end{figure}

We present in the following just one case, giving the results in Fig.\ref{fig4}, for 
$H_i=10 m$ and the parameter values $\alpha=0.0085$ and $\chi_0=0.5M_{pl}$.
In this specific case one can see, in the left plot of Fig.\ref{fig4}, that the quantum 
backreaction on the effective expansion scalar is positive at the beginning of the inflationary 
period and diminishes monotonically with the effective number of e-foldings 
$N_{eff}=\int dt H_{eff}$ up to the point where it has become negative enough to stop 
the observed accelerated expansion earlier than what would occur in the absence of
backreaction. For the parameters chosen in this simulation this happens at 
$N_{eff}\simeq 89$ compared to the case without backreaction when $N=148.5$.
Here, the end point of inflation is taken to correspond to $H(t_f)=m$. Therefore, the 
observer comoving with the moduli field is experiencing an apparently shorter inflationary period.

The effective equation of state is affected in the following way: $w_{eff}$ receives a positive 
backreaction at the beginning which leads it farther away from a de Sitter like equation of state, 
but during the evolution the difference decreases. As shown in the figure, the background 
value is always more de Sitter like. The observer experiences a kind of ``apparent inverted 
chaotic inflation (AICI)", in the sense that the evolution of the effective equation of state 
is somehow reversed.

\section{Conclusions}

In this work we focused  on the backreaction of long wavelength quantum fluctuations 
produced during inflation. We have adopted a computational scheme, recently 
developed~\cite{GMV1,GMV2}, which is covariant and has guided us to define gauge 
invariant observables in the sense that any observer, in whatever reference system, 
agrees on them.

Since in late time cosmology  time is measured in terms of the temperature of the CMB, 
which is subdominant at later times compared to other matter, one might try to measure, 
even during an inflationary regime, an effective expansion rate in terms of a clock associated 
to a test field which is not directly affecting the background geometry. This measure 
nevertheless may be affected by strong quantum fluctuations of the gravity-matter system 
which must be taken into account in a gauge invariant way.

Taking advantage of the results obtained in \cite{FMSVV2}, for a chaotic inflationary model 
with the presence of a test field, and using a stochastic approach in the slow-roll regime, 
we have thus studied a particular class of observables which are related to measurements 
performed by an observer comoving with a test field. In particular, we have analyzed 
the average effective expansion rate and the effective equation of state as seen by such 
class of observers. We find that - except for a short period after the onset of inflation -
the long wavelength fluctuations lead to a reduction of the expansion rate of space
compared to what an observer in an unperturbed space-time would see. This leads
to a reduction in the measured length of the inflationary period.
Throughout our work we have taken into account the constraints on the parameter
space of the model in which the dynamics can be studied at second order in perturbation 
theory in a reliable way.

As we have shown explicitly, then during the final stages of the inflationary phase, and
in the case of a sufficiently light test field, the quantum corrections of the average
expansion rate, as measured by our comoving field, are dominated by the term proportional 
to the inverse of the test field mass to the fourth power times the square of the homogeneous 
initial value  of the test field.  As a consequence, for an extremely light test field, the
backreaction may even overcome the value given by the classical dynamics, even if  
cosmological perturbation theory for the field fluctuations and the test field approximation 
are both valid. In such a case one should employ a non perturbative approach to study 
the observables. One may argue that in such a regime these observables are very peculiar. 
In a sense they feel the quantum effects in an amplified way.

We have investigated the region of parameter space where the quantum backreaction 
effects are of the same order of magnitude as the background quantities and we have shown 
that these effects can stop the "observed" inflation much earlier. We have also noted that the measured 
effective equation of state has an inverted time evolution starting farther from but 
getting closer to a de Sitter like phase with the evolution. 
We have called this kind of observed behavior ``apparent inverted chaotic inflation (AICI)".
Let us point out that this ``apparent" dynamics cannot be mimicked by another chaotic 
inflationary model (even with an higher mass) since the effective equation of state has a 
completely different "inverted" dynamics.

A way to give a physical explanation for such possibly non-perturbative effects of 
backreaction, as seen by observers comoving with the test field $\chi$, comes from 
looking at the relative gauge transformation that connects a general gauge to the 
gauge U$\chi$FG. Starting from Eqs.(\ref{GT1},\ref{GT2}) it is easy to see as such a gauge 
transformation corresponds to a big time shift (from Eq.(\ref{311})) when 
$\dot{\chi}^{(0)}\ll 1$, and a non-perturbative effect of the backreaction is possible.
Namely, we find that non-perturbative backreaction corresponds to a gauge trasformation 
which is no longer infinitesimal.

Note also that we have made a test field approximation for the scalar $\chi$ associated 
to the observer measuring $H_{eff}$. This means that the dynamics (solution of the 
semiclassical equations of motion) of the inflaton and of the metric is not affected
by the presence of $\chi$. This fact helps to understand that the results found for the 
observables considered here are therefore non related to an absolute concept of quantum 
backreaction on the dynamics of the expansion of the Universe, which is probably a 
very difficult concept to define. It is related instead to observations which the specific 
class of dynamical observers can make.

Summarizing our findings, from our analysis we therefore deduce that a gedanken experiment 
with observers of the kind discussed here may be useful to extract "amplified" quantum 
corrections in the associated observables. Insisting on considering such observables as 
physically relevant, one faces in general a computability problem.
If the backreaction on the GI observable associated to the effective expansion rate is small 
then one can proceed with our perturbative approach which stops at second order and 
obtain a reasonable estimate. We have found this is possible only in a bounded parameter 
region. When outside of this region, the perturbative analysis for the observer level 
could fail. We lack other tools to perform a different investigation.
Therefore in such a case we are not sure that the results found can be extrapolated, at least 
qualitatively, to a non-perturbative region.

\vskip 0.3cm
\noindent
{\bf Acknowledgements}

We wish to thank Gabriele Veneziano for stimulating discussions.
GPV thanks the Cosmology Group of the University of Geneva for its kind hospitality.
GM is supported by the Marie Curie IEF, Project NeBRiC - ``Non-linear effects and backreaction 
in classical and quantum cosmology". The research of RB is supported by an NSERC Discovery 
Grant and by funds from the Canada Research Chair program.


\end{document}